\documentclass[preprint,
showkeys,
amssymb,aps,prb]{revtex4}

\usepackage{graphicx}
\usepackage{bm}

\begin{document}



\title{Oscillatory and Vanishing Resistance States \\
in Microwave Irradiated 2D Electron Systems}

\author{R. R. Du\footnote{Corresponding author: rrd@physics.utah.edu}} 
\author{M. A. Zudov\footnote{Present address: School of Physics and Astronomy, 
University of Minnesota, Minneapolis, MN 55455, USA}}
\author{M. A. Zudov}
\author{C. L. Yang}
\author{Z. Q. Yuan}
\affiliation{Department of Physics, University of Utah, Salt Lake 
City, Utah 84112, USA}

\author{L. N. Pfeiffer}
\author{K. W. West}
\affiliation{Bell Laboratories, Lucent Technologies, 
New Jersey 07974, USA}  

\begin{abstract}
Giant-amplitude oscillations in dc magnetoresistance of a high-mobility 
two-dimensional electron system can be induced by millimeterwave 
irradiations, leading to zero-resistance states at the oscillation minima.   
Following a brief overview of the now well-known phenomenon, this paper reports on 
aspects of more recent experiments on the subject. These are: new 
zero-resistance states associated with multi-photon processes;
suppression of Shubnikov-de Haas oscillations by  high-frequency microwaves; 
and microwave photoconductivity  of a high-mobility two-dimensional hole system.
\end{abstract}


\maketitle

\section{Introduction}
Under irradiation of microwaves dramatic new effects occur 
in two-dimensional electron systems (2DES) in GaAs-AlGaAs 
heterostructures previously used to study quantum Hall 
effects (QHE).\cite{r1} As first reported by Zudov {\it et al.},\cite{r2}
and also by Ye {\it et al.},\cite{r3} in the millimeterwave (MW) frequency range 
($f = 30 - 150$ GHz) and in a small magnetic field 
($B < 0.5$~T), at low temperatures ($T\sim1$ K) a new type of oscillations 
arises in the magnetoresistance $R_{xx}$
of a high-mobility 2DES. These oscillations are periodic  
in $1/B$ and can occur in a $B$ weaker than the onset of  
Shubinikov-de Haas oscillations (SdH). Characteristically, their 
period is controlled by the ratio of the microwave frequency 
to the electron cyclotron frequency, $\epsilon = \omega/\omega_{c}$, 
where $\omega_c= eB/m^*$, $m^*$ is the effective 
mass of the conduction band electrons in GaAs. 
In particular, the oscillations can be associated with cyclotron 
resonance (CR, $\epsilon =1$) and its harmonics ($\epsilon = 2, 3, 4, \dots$). 
Although the photoconductivity (PC) effect was 
well-known in the literature\cite{r4,r5,r6} such findings came as 
a surprise. Not only the observation of higher-order CR  
(hence oscillations rather than a single peak), but also its 
large amplitude, was completely unexpected. Moreover, later it was found 
that in samples of even higher mobility, the oscillation minima approach zero. 
Recently, Mani {\it et al.}\cite{r7} and Zudov {\it et al.}\cite{r8}
reported the observation of new ``zero-resistance states'' (ZRS) associated with such 
minima in very clean samples. At low temperatures, ZRS are characterized 
by an exponentially vanishing diagonal resistance and an 
essentially classical Hall resistance .  
Such a new effect in a 2DES induced by microwaves is a subject of much 
current experimental\cite{r9,r10,r11,r12,r13,r14} and theoretical 
work.\cite{r15,r16,r17,r18,r19,r20,r21} Some of the theoretical ideas 
can be traced back to earlier studies of negative photoconductivity 
phenomena in semiconductors.\cite{r22} 
These models, as well as others, remain to be tested experimentally. The observation 
of ZRS is complemented by an observation of zero-conductance 
states under similar experimental conditions, but in 
Corbino samples.\cite{r9}

\section{Overview of the Effect}
Photoconductivity related to CR of a 2DES was first reported by
Maan {\it et al.}\cite{r4} in the far-infrared 
frequency range and in a strong magnetic field.  In the MW frequency range, 
distinct magnetoresistance signals due to electron spin 
resonance (Dobers {\it et al.}\cite{r5}) or magnetoplasmon resonance
 (Vasiliaou {\it et al.}\cite{r6}) were observed in 
2DES in GaAs-AlGaAs heterostructures. 
In these earlier studies using samples of electron mobility 
$\mu \sim 1 \times 10^6$ cm$^2$/Vs or less, the PC signal exhibits a
single peak in the $R_{xx}$, and its amplitude is typically less than 
1\%. The MW-induced giant amplitude magnetorsistance oscillations 
were discovered\cite{r2} using a high-mobility ($3 \times 10^6$ cm$^2$/Vs) 
2DES. In a small magnetic field the $R_{xx}$ shows periodic (in $1/B$) 
oscillations, with the $B$ positions of major maxima and minima 
in the oscillatory structure roughly conforming to\cite{r12} 
\begin{equation}
\omega/\omega_c=\epsilon= \left\{
\begin{array}{ll}
j & \rm{\ \ maxima\ \ \ }\\ j+1/2 & \rm{\ \ minima\ \ \ } \end{array}
\right. j = 1,\ 2,\ 3,\ \dots .  
\end{equation}
Phenomenologically such oscillations resemble 
SdH oscillations except that their period relates to $\epsilon$
rather than to $\nu$, the Landau level filling factor. Most 
remarkably, the $R_{xx}$ at the minima was found 
to reduce from its dark value, indicating  
of negative PC. This was a crucial step leading to the discovery 
of ZRS; increasing sample mobility favors oscillation amplitude, 
and, therefore, the $R_{xx}$ value at the minima becomes progressively 
smaller under similar experimental conditions.
\begin{figure}
\includegraphics[width=8cm]{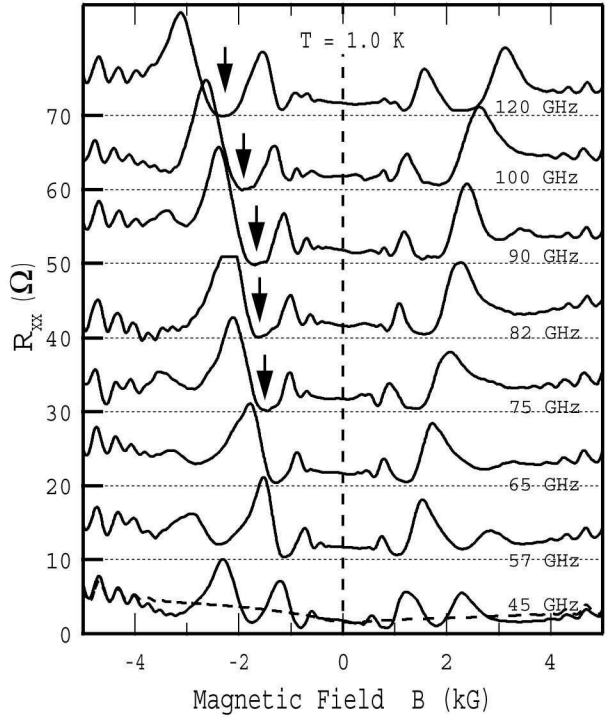}
\caption{Microwave induced oscillation minimum evolves into zero-resistance 
state as the microwave frequency is increased. The sample is a 
GaAs/Al$_{0.3}$Ga$_{0.7}$As heterojunction, having a 2DES of density
$n_e=2.7\times10^{11}$ cm$^{-2}$ and mobility $\mu=1 \times10^7$ cm$^2$/Vs.
}
\label{fig1}
\end{figure}

Vanishingly small $R_{xx}$ at the minima was initially observed 
in a GaAs/Al$_{0.3}$Ga$_{0.7}$As heterojunction 2DES approaching a mobility 
$\mu=10 \times10^6$ cm$^2$/Vs.\cite{r23} The data of Yang {\it et al.} 
are shown in Fig.~\ref{fig1}. 
As marked by arrows, in this sample the strongest minimum 
in the $R_{xx}$ evolves into ZRS as the MW frequency is increased 
(hence, the minimum moves towards higher $B$).
 
\begin{figure}
\includegraphics[width=10cm]{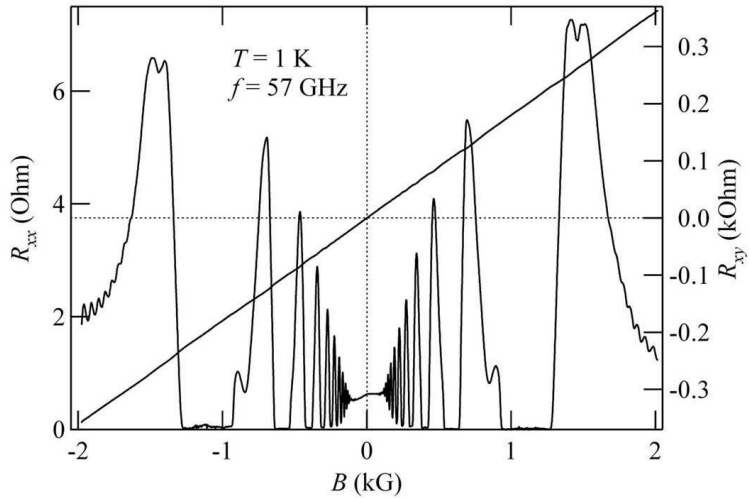}
\caption{Electrical transport data in a Hall sample of a very high 
mobility quantum well exposed to millimeter waves. The magnetoresistance 
$R_{xx}$ shows sharp oscillations and a series of zero-resistance states; Hall 
resistance $R_{xy}$ shows an essentially classical, linear 
dependence on B.
}
\label{fig2}
\end{figure}

Our current data from an ultraclean 2DES\cite{r8} are shown in Fig.~\ref{fig2}. 
The sample was cleaved from a Al$_{0.24}$Ga$_{0.76}$As/GaAs/Al$_{0.24}$Ga$_{0.76}$As 
QW (width 30 nm), with a mobility $\mu=25\times10^6$ cm$^2$/Vs and a density
$n_e=3.5\times10^{11}$ cm$^{-2}$, respectively. With the irradiation of MW 
($f = 57$ GHz and the power incident on the sample surface $P\sim 100\ \mu$W), 
the $R_{xx}$ shows a series of oscillations and 
the novel zero-resistance states at the first four minima. 

The width of the ZRS encompasses a large range of the filling 
factor $\nu$. For example, the first ZRS spans from $\nu\sim 115$ 
to $\nu\sim155$ and similar 
width of $\Delta\nu\sim40$ is observed for other 
ZRS minima in this sample. These data indicate that the transport 
under MWs is controlled by $\epsilon$ rather than by $\nu$. 
While the $R_{xx}$ here remarkably resembles that in 
QHE, the Hall resistance, $R_{xy}$, shows no sign of forming
plateaus. In 
the very-high mobility 2DES,  the major peaks of MW-induced oscillations
(which are relevant to ZRS regime) can still be well described 
by Eq. 1 for $j = 1, 2, 3$,  but their shapes are largely asymmetric. 
Higher-order peaks are better described by a $B$-dependent phase shift 
with respect to CR harmonics ($j = 4, 5, \dots$ ), which saturates at $-1/4$.\cite{r12}
\begin{figure}
\includegraphics[width=10cm]{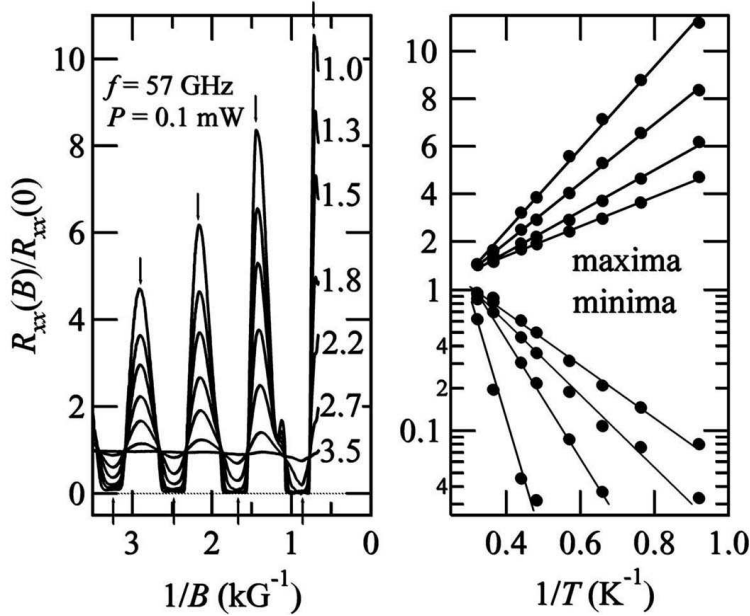}
\caption{The temperature dependence of $R_{xx}$ peaks and 
ZRS minima, showing a roughly linear dependence of the peak values 
on inverse temperature, and a thermally activated resistance 
at the minima.
}
\end{figure}

One of the most interesting and puzzling results in the ZRS regime 
is the temperature dependence. The $1/B$ oscillations are shown 
in Fig.~3 for different temperatures but under the same MW irradiation. 
Below a temperature $T\sim 4$ K both the maxima and 
the minima exhibit strong temperature dependence. Using standard 
Arrhenius plot we present the $R_{xx}(B)/R_{xx}(0)$ 
of the first four minima on a logarithmic scale, versus $1/T$. 
All four minima conform to a general expression, $R_{xx}(T)\sim exp(-T_0/T)$,
over at least one decade in the $R_{xx}$. 
Unusually large $T_0$ is found for these minima. For example, $T_0\approx20$ K 
for the first minimum, exceeds both the MW photon 
energy ($\sim 3$ K) and the Landau level spacing ($\sim2$ K) by about an order of 
magnitude. As also shown in Fig. 3, the $R_{xx}$ peaks are roughly linear with $1/T$. 

Using Corbino samples, we have observed zero-conductance states (ZCS) 
under similar experimental conditions.\cite{r9} Typical
data from a high mobility ($\mu\approx12.8\times10^{11}$ cm$^2$/Vs)  
2DEG Corbino ring are shown in Fig.~4.
Temperature dependence, as well as I--V measurements at ZCS,  shows that 2DES 
behaves like a macroscopic insulator in this regime.\cite{r9}
\begin{figure}
\includegraphics[width=8cm]{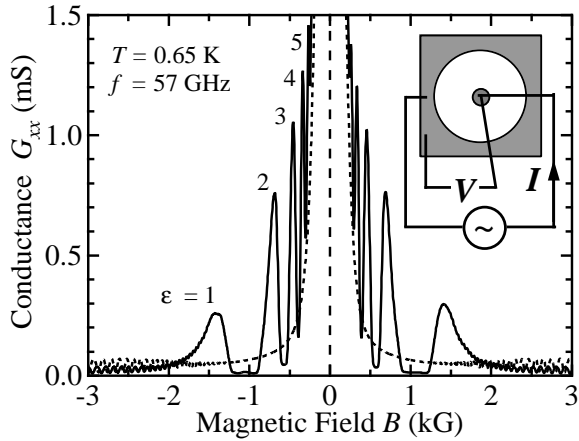}
\caption{Microwave-induced conductance oscillations and zero-conductance 
states in 2DES measured in a Corbino sample.
}
\end{figure}

\section{Multi-Photon Processes}
Remarkably, we have observed a series of new ZRS associated with fractional 
$\epsilon$, such as 
\begin{figure}
\includegraphics[width=6.5cm]{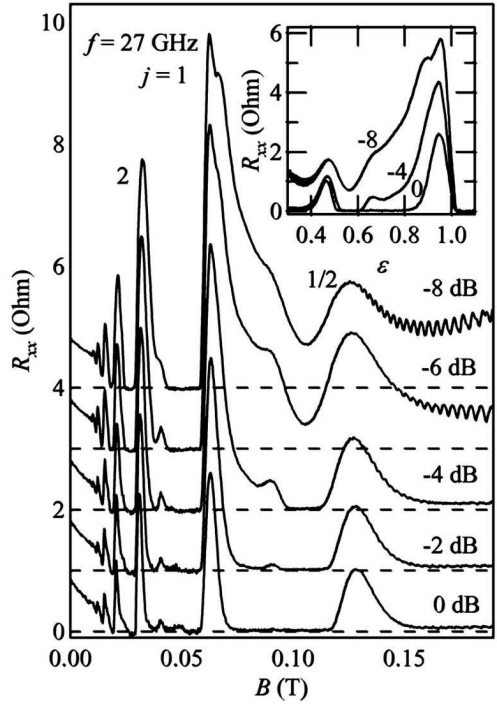}
\caption{Formation of zero-resistance states associated with  
transitions promoted by two microwave photons is shown. Maxima appear near 
$\epsilon=1/2$ and $\epsilon=3/2$; the associated minima develop into 
ZRS as the microwave power is increased 
(attenuation of the microwave power is marked next to the $R_{xx}$ traces).   
}
\end{figure}
$\epsilon=1/2$ and $3/2$, in an ultraclean 2DEG sample (similar 
to the sample used in Fig.~2) exposed to MW of lower frequency ($f<30$~GHz). 
Figure~5 shows 
the development of such ZRS with increasing
MW power $P$.
At all $P$ values the previously reported ``integer'' 
ZRS are observed. Beyond the $\epsilon = 1$ peak, a $\epsilon = 1/2$ peak and associated 
minimum (around $B\sim0.1$ T) are already visible at 
lower power (attenuation $-8$~dB). With increasing $P$, 
this minimum develops into a new wide ZRS. The $\epsilon=1/2$ 
structure has first been noticed in Ref.~\onlinecite{r2} and  can also be seen in Fig.~1;
the feature becomes stronger in samples with higher mobility.\cite{r8} 
Dorozhkin\cite{r10} and Willett {\it et al.}\cite{r11} have also reported 
features associated with $\epsilon =1/2$. In this experiment we also observe  
fractional ZRS forming about $\epsilon=3/2$. Another important observation is the 
(overall) suppression of the $R_{xx}$ at $\epsilon<1/2$.
\begin{figure}
\includegraphics[width=7cm]{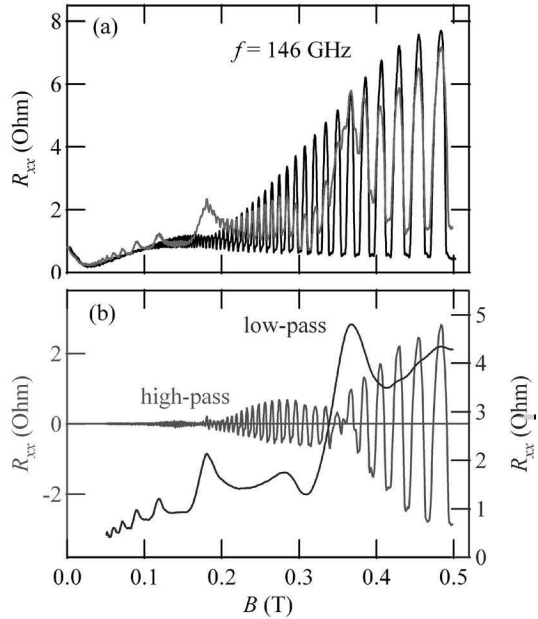}
\caption{The Shubinikov de-Haas oscillations are modulated by 
microwave-induced oscillations. 
}
\end{figure}

From the $B$ position we interpret these new ZRS as 
resulting from two-photon processes, where two photons of frequency $\omega$ 
participate in the transition between Landau levels separated 
by $\omega_{c}$ and $3\omega_c$, respectively\cite{r12}. This observation 
may suggest the importance of virtual processes in 
the microwave-induced oscillations and ZRS. 

In the higher frequency region ($>100$ GHz), we observed a strong modulation of SdH 
oscillations by the presence of MW. This phenomenon has been reported 
in Ref.~\onlinecite{r13} for samples of a moderately-high mobility. The data in 
Fig.~6 shows similar results from an ultra-clean 2DES. In addition to the
first node, higher-order nodes were observed in this sample. 

\section{Photoconductivity of a 2D Hole System}
It is interesting to investigate the photoconductivity oscillations 
and ZRS in a high-mobility 2D hole system (2DHS) primarily for the 
following reasons: 1) comparing with 2D electron system, 2DHS has different 
band structure parameters such as larger effective mass and 
significant nonparabolicity of the dispersion; 2) 2DHS 
possesses strong spin-orbit interactions. Our preliminary result for 
a 2DHS is shown in Fig.~7. The 2DHS is from a (001) 
GaAs/Al$_{0.4}$Ga$_{0.6}$As 
QW grown by MBE. The QW has a well width of 15 nm; holes are 
provided by a C delta-doping layer situated 50 nm away from the 
well. A 100-$\mu$m-wide Hall bar, on which measurements were performed, 
is defined by lithography and is contacted by In/Zn alloy. 
\begin{figure}[b]
\includegraphics[width=10cm]{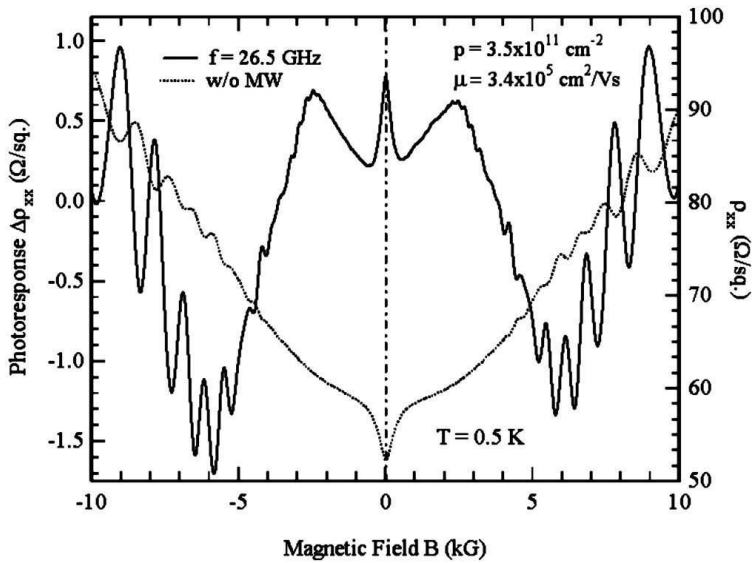}
\caption{A high-mobility 2D hole system in GaAs/AlGaAs quantum well 
shows sharp positive peak and negative valley in photoresistance $\Delta\rho_{xx}$ 
under microwave irradiation. 
}
\end{figure}

To simultaneously detect photoresistivity, $\Delta\rho_{xx}= \rho_{xx}^{MW}
- \rho_{xx}^{DARK}$, and resistivity, $\rho_{xx}$, a double-modulation 
technique was employed. The main feature 
of $\Delta\rho_{xx}$, as shown for MW frequency of 26.5 
GHz, is a positive peak at $B\approx 2.5$ kG and a negative 
valley (a suppression of  $\rho_{xx}$) around $\sim6$ kG, both 
having relatively small magnitude. The 2.5 kG peak could be associated 
with the CR of the 2D holes which have an effective mass $m^*\sim0.3\ m_e$ 
(measured in separate transmission experiments in a 
similar $B$ range; for measurement techniques, see Ref.~\onlinecite{r24}). 
The nature of the 6 kG valley is not yet clear. 

From the mobility of the holes, $\mu = 3.4\times10^{5}$ cm$^2$/Vs 
and a mass value of 0.3~m$_{e}$, we estimate a transport scattering time 
of  $\tau_{t} = \mu m^*/e \approx 58$ ps. This value is a factor of two smaller 
than that of the 2DES (where $\tau_t\approx 115$ ps) in 
which the giant amplitude oscillations were 
originally observed,\cite{r2} a fact that may be considered in explaining the lack of 
oscillations in this 2DHS. However, the band structure of holes must be taken into account
to understand these data.

\section{Summary}

Oscillatory and vanishing magnetoresistance can be induced by 
millimeterwave irradiation in high mobility two-dimensional 
electronic systems traditionally used for studies of the quantum 
Hall effect. Depending on the sample geometry, zero resistance 
states (in Hall samples) or zero conductance states (in Corbino 
samples) are observed. Some new aspects of the observations are 
reported here which may help to distinguish between competing 
theoretical models of the phenomenon.


\begin{acknowledgements}
This work was supported by NSF, by DARPA, and by A. P. Sloan 
Foundation. 
\end{acknowledgements}

\end{document}